\documentclass[conference]{IEEEtran}
\IEEEoverridecommandlockouts
\usepackage{cite}
\usepackage{amsmath,amssymb,amsfonts}
\usepackage{graphicx}
\usepackage{xcolor}
\usepackage{booktabs}
\usepackage{multirow}

\begin{document}
\title{Beyond MACs: Deployment-Constrained Design Space Exploration for Tensorized Neural Networks}
\author{%
\IEEEauthorblockN{
Jinsong Zhang\textsuperscript{1,*},
Minghe Li\textsuperscript{1,2,*},
Jiayi Tian\textsuperscript{1},
Jinming Lu\textsuperscript{1,\dag},
Zheng Zhang\textsuperscript{1,\dag}
}
\IEEEauthorblockA{
Email: \{jinsongzhang, jiayi\_tian\}@ucsb.edu;
zhengzhang@ece.ucsb.edu\\
mli247@jh.edu;
jinmingmaster@gmail.com\\
\textsuperscript{1}Department of Electrical and Computer Engineering,
University of California, Santa Barbara, CA, USA\\
\textsuperscript{2}Department of Electrical and Computer Engineering, Johns Hopkins University,
Baltimore, MD, USA
}
\thanks{
\textsuperscript{*}Equal contribution.
\textsuperscript{\dag}Corresponding authors.
}
}
\maketitle

\begin{abstract}
Tensor-train (TT) compression reduces neural-network parameters, but the latency of a tensorized layer is set by its contraction order and its mapping onto hardware: a contraction path with fewer multiply--accumulate operations (MACs) is not necessarily faster on a systolic array. We present a deployment-constrained, latency-guided design space exploration (DSE) framework for fixed-rank tensorized neural networks (TNNs) under a fixed processing-element (PE) budget. Using MAC-guided path pruning, it evaluates feasible combinations of contraction paths, three PE-array partitions, and input-, output-, and weight-stationary dataflows. By minimizing the sum of inference and training latency estimates, it selects a single physical PE partition shared across layers and both modes, while allowing layer- and mode-specific paths and dataflows. On ResNet-18 and ViT-Ti/4, 25\%--50\% of the layers select a path other than the minimum-MAC one, the best dataflow varies across layers and models, and on a representative ViT layer the latency-selected candidate reduces simulated latency by 23.5\% relative to the minimum-MAC candidate. Deployed on a Xilinx VU9P FPGA accelerator that supports both inference and training, the complete TT+INT8 system achieves 3.28$\times$--4.00$\times$ inference and 3.42$\times$--3.85$\times$ training speedups over dense implementations with the same PE budget.
\end{abstract}

\section{Introduction}
\begin{figure}[!t]
\centering
\includegraphics[width=\linewidth]{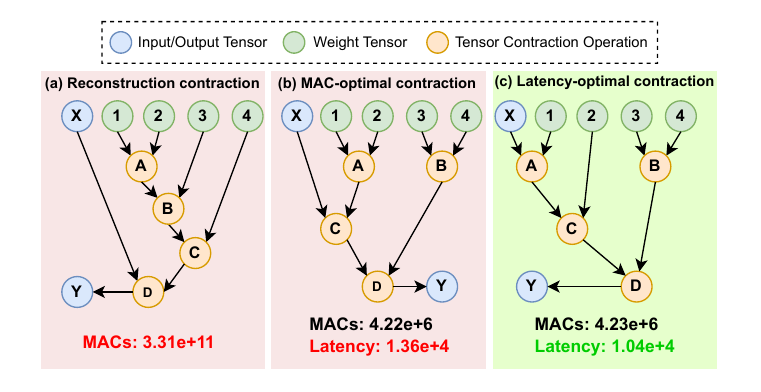}
\caption{Contraction paths of a TT layer in ViT-Ti/4 (CIFAR-10): (a) reconstruction-based, (b) minimum-MAC ($4.22\times10^6$ MACs, $1.36\times10^4$ cycles), and (c) latency-selected ($4.23\times10^6$ MACs, $1.04\times10^4$ cycles).}
\label{fig:path}
\end{figure}

Deep neural networks (DNNs) have advanced image classification~\cite{he2016deep} and object detection~\cite{szegedy2013deep}, but their computation and storage demands challenge edge hardware. Tensor decomposition~\cite{kolda2009tensor,sidiropoulos2017tensor}, and tensor-train (TT) decomposition in particular, represents neural-network weights with compact low-rank cores, reducing the storage cost of edge deployment~\cite{novikov2015tensorizing,oseledets2011tensor,yang2024comera}. Related low-rank and tensor-compression work covers convolutional, recurrent, and large language models as well as rank selection~\cite{kim2016compression,tjandra2017compressing,yin2021towardsrnn,yin2021towardsdnn,yin2022hodec,tian2025flat,hawkins2021bayesian,hawkins2022towards}. Quantization~\cite{wang2022niti,lu2022eta,xu2022mandheling}, sparsity~\cite{wang2019low,lu2022theta,fang2023efficient,fang2023cest}, memory-efficient on-device learning~\cite{wang2022melon}, and quantization- or hardware-aware compression~\cite{yang2023quantization,xiao2023haloc} further reduce deployment cost. However, fewer parameters or multiply--accumulate operations (MACs) do not necessarily imply lower hardware latency. A TT layer executes a sequence of contractions, and their order changes intermediate matrix shapes, processing-element (PE) utilization, and operand reuse. In Fig.~\ref{fig:path}, a latency-selected configuration for a ViT layer uses approximately 0.24\% more MACs than the minimum-MAC candidate yet reduces simulated latency by 23.5\%.

Mapping decisions have different deployment granularity. Contraction paths and input-, output-, or weight-stationary (IS/OS/WS) dataflows can vary by layer and execution mode, whereas the physical PE-array partition remains fixed in the FPGA. Independently choosing each layer's preferred partition is therefore not deployable; fixing all mappings instead ignores heterogeneous contraction shapes and opportunities for parallel branches.

\textbf{Relation to prior work.}
Prior studies have demonstrated the benefits of hardware-aware tensor decomposition and contraction mapping. HEAT explores tensorization shape and rank alongside contraction paths and fused Einsum mapping~\cite{gu2022heat}. TetriX combines mapping-aware sequence selection with flexible hardware and tensor-layout management~\cite{zhang2024tetrix}. FETTA searches contraction sequences and employs reconfigurable engines for TNN training~\cite{lu2025fetta}, while related FPGA work uses bidirectional TT contractions~\cite{tian2025ultra} and ultra-low-memory tensorized training~\cite{chen20213u}. Systolic arrays~\cite{kung1982systolic} with weight-, output-, or input-stationary dataflows~\cite{chen2016eyeriss,jouppi2017tpu} are widely used for DNN acceleration, and tools such as Timeloop, MAESTRO, and SCALE-Sim explore their mappings~\cite{parashar2019timeloop,kwon2019maestro,raj2025scale}. These tools take the operator shapes as given, whereas in a TT layer the contraction order itself determines these shapes. Complementary FPGA training accelerators investigate data reshaping and low-precision arithmetic~\cite{tang2022ef,guo2023boost}. We focus on a deployment-level optimization challenge: a distinct deployment constraint where \emph{one fixed PE partition must serve all layers and both inference and training of a given fixed-rank TNN}, while paths and dataflows remain layer- and mode-specific.

To address this deployment constraint, we develop a latency-guided design space exploration (DSE) framework that jointly considers contraction paths, PE-array partitions, and stationary dataflows. For each candidate partition, layer-specific inference and training latency estimates are minimized and summed to choose one shared partition. The selected mappings run on an FPGA GEMM and streaming TT engine that supports both inference and training.

Our contributions are:
\begin{itemize}
 \item \textbf{Deployment-constrained DSE.} Under a fixed PE budget, we select one partition for inference and training jointly, with layer- and mode-specific paths and dataflows, minimizing modeled latency over retained feasible candidates.
 \item \textbf{Hardware-sensitive mapping analysis.} In 25\%--50\% of evaluated layers, the selected path is not minimum-MAC; one ViT layer exhibits 23.5\% lower simulated latency. Dataflow and split-array preferences also vary across workloads.
 \item \textbf{FPGA validation.} A VU9P implementation supports CNN and ViT inference and training and achieves 3.28$\times$--4.00$\times$ inference and 3.42$\times$--3.85$\times$ training speedups over matched INT8 dense implementations.
\end{itemize}

\section{Background: Tensorized Layers}

\begin{figure}[!t]
\centering
\includegraphics[width=0.9\columnwidth]{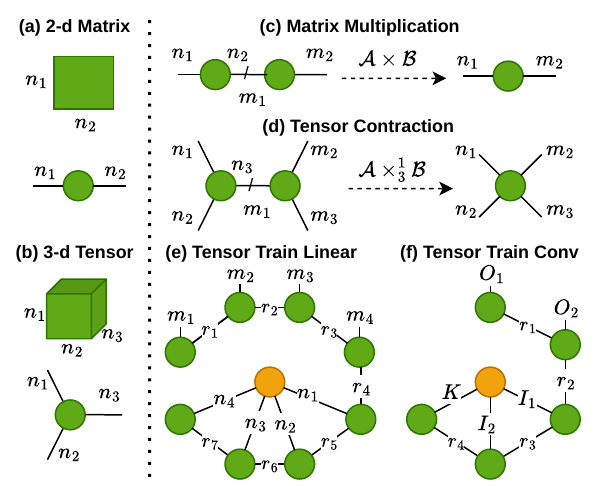}
\caption{Tensor graphs: (a) matrix, (b) tensor,
(c) matrix multiplication, (d) contraction,
(e) TT linear layer, and (f) TT convolution layer.}
\label{fig:tensor-basic}
\end{figure}

A $d$-way tensor $\mathcal A\in\mathbb R^{n_1\times\cdots\times n_d}$ has $d$ modes of sizes $n_1,\ldots,n_d$~\cite{kolda2009tensor}. In a tensor graph (Fig.~\ref{fig:tensor-basic}(a)--(b)), a $d$-way tensor is a node with $d$ edges, so a matrix is a node with two edges. Contracting mode $s$ of $\mathcal A$ with mode $t$ of $\mathcal B\in\mathbb R^{m_1\times\cdots\times m_q}$ ($n_s=m_t$) is written as $\mathcal C=\mathcal A\times_s^t\mathcal B$, which sums over the shared mode; $\mathcal C$ keeps the remaining modes of $\mathcal A$ followed by those of $\mathcal B$. Fig.~\ref{fig:tensor-basic}(c)--(d) shows contractions of 2-way and 3-way tensors, the former being a matrix multiplication; the free (unconnected) edges give the modes of the result. After reshaping, every contraction is executed as a GEMM, and different contraction orders produce different intermediate shapes and MAC counts.

\textbf{TT linear layer.}
For $\mathbf W\in\mathbb R^{M\times N}$ with $M=\prod_{i=1}^d m_i$ and $N=\prod_{i=1}^d n_i$, $\mathbf W$ is reshaped into a $2d$-way tensor $\mathcal W$ of size $m_1\times\cdots\times m_d\times n_1\times\cdots\times n_d$ and represented by $2d$ TT cores~\cite{oseledets2011tensor} as $\mathcal W=\mathcal G_1\times_3^1\mathcal G_2\times_3^1\cdots\times_3^1\mathcal G_{2d}$, where $\mathcal G_k\in\mathbb R^{r_{k-1}\times q_k\times r_k}$, $q_k=m_k$ for $k\le d$, $q_k=n_{k-d}$ otherwise, and $r_0=r_{2d}=1$. The cores hold $\sum_k r_{k-1}q_kr_k$ parameters instead of $MN$. The forward pass maps $\mathcal X\in\mathbb R^{n_1\times\cdots\times n_d}$ to $\mathcal Y\in\mathbb R^{m_1\times\cdots\times m_d}$. Let $\mathbf i=(i_1,\ldots,i_d)$ and $\mathbf j=(j_1,\ldots,j_d)$. Then
\begin{equation}
\label{eq:tt-fwd}
\mathcal Y[\mathbf i]=\sum_{\mathbf j}
\Bigl(\prod_{k=1}^{d}\mathcal G_k[i_k]\Bigr)
\Bigl(\prod_{k=1}^{d}\mathcal G_{d+k}[j_k]\Bigr)
\mathcal X[\mathbf j],
\end{equation}
where $\mathcal G_k[i]$ is the $i$-th $r_{k-1}\times r_k$ slice of $\mathcal G_k$ (Fig.~\ref{fig:tensor-basic}(e)). The cores and input admit multiple valid contraction orders that preserve the intended tensor-network contraction.

\textbf{TT convolution layer.}
For $\mathcal W\in\mathbb R^{C_{out}\times C_{in}\times K_h\times K_w}$, we factor $C_{out}=O_1O_2$ and $C_{in}=I_1I_2$ and set $K=K_hK_w$, giving five cores with mode sizes $(O_1,O_2,I_1,I_2,K)$. This reduces the parameter count from $O(C_{out}C_{in}K)$ to $O(r^2(C_{out}+C_{in})+rK)$. For the unfolded input $\mathcal X_{\mathrm{unf}}\in\mathbb R^{I_1\times I_2\times K\times N_{\mathrm p}}$ with $N_{\mathrm p}$ spatial patches indexed by $\nu$, let $\mathbf o=(o_1,o_2)$, $\mathbf u=(i_1,i_2,k)$, and $\mathbf s=(o_1,o_2,i_1,i_2,k)$. Then
\begin{equation}
\label{eq:tt-conv}
\mathcal Y_{\mathrm{col}}[\mathbf o,\nu]
=\sum_{\mathbf u}
\Bigl(\prod_{t=1}^{5}\mathcal G_t[s_t]\Bigr)
\mathcal X_{\mathrm{unf}}[\mathbf u,\nu],
\end{equation}
so the input interacts with the input-side cores $\mathcal G_3$--$\mathcal G_5$, and the output-side cores $\mathcal G_1,\mathcal G_2$ generate the output channels (Fig.~\ref{fig:tensor-basic}(f)). In training, the backward pass adds contractions for the input and core gradients, which are also free to reorder.

\begin{figure*}[t]
\centering
\includegraphics[width=0.95\linewidth]{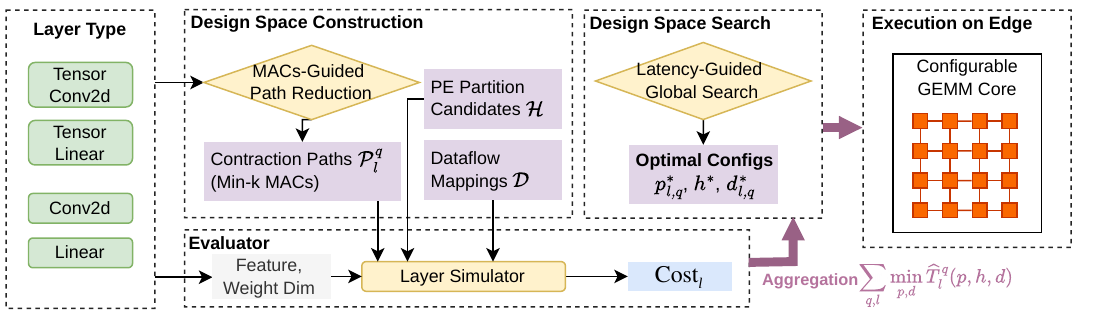}
\caption{Deployment-constrained DSE flow. MAC-guided pruning generates candidate paths, and a simulator evaluates feasible path/partition/dataflow combinations. Layer-level inference and training latency estimates are aggregated for each partition; the partition with the lowest combined model latency is fixed for deployment. Paths and dataflows can differ by layer and execution mode.}
\label{fig:framework}
\end{figure*}

\section{Deployment-Constrained DSE}
\label{sec:framework}
Given a fixed-rank TNN and a fixed PE budget, the proposed DSE framework considers three physical array partitions (Fig.~\ref{fig:framework}). A partition is selected once per model and is shared by inference and training; the executed contraction paths and dataflows may depend on both the layer and the execution mode. The training accelerator extends the inference datapath with backward and weight-update support, so both modes execute on the same PE-array organization and the partition must serve both.

\subsection{Problem Formulation}
Let $\mathcal Q=\{\mathrm{inf},\mathrm{train}\}$ denote the two execution modes and $l=1,\ldots,L$ index the model layers. The TT ranks, model weights, and INT8 arithmetic are fixed before the mapping search. For layer $l$ and mode $q$, let $\mathcal P_l^{q}$ denote at most $K$ candidate contraction paths, and let $\mathcal D=\{\mathrm{IS},\mathrm{OS},\mathrm{WS}\}$ denote the available stationary dataflows. The global partition $h$ is chosen from
$\mathcal H=\{1\!\times\!1,1\!\times\!2,2\!\times\!1\}$; the last two configurations split the same PE budget into two equal subarrays.
The feasible path/dataflow pairs under $h$ form $\mathcal F_l^q(h)\subseteq\mathcal P_l^q\times\mathcal D$. Denote the simulated latency of a feasible mapping by $\widehat T_l^q(p,h,d)$. We select
\begin{equation}
\label{eq:obj}
h^*=\underset{h\in\mathcal H}{\arg\min}\;\sum_{q\in\mathcal Q}\sum_{l=1}^{L}
\min_{(p,d)\in\mathcal F_l^q(h)}\widehat T_l^q(p,h,d).
\end{equation}
For each candidate $h$, the inner minimizations select the path and dataflow for each layer and mode; their inference and training costs are summed with equal weights to select one shared $h^*$. After selecting $h^*$, each $(p_{l,q}^*,d_{l,q}^*)$ is the corresponding inner minimizer. A partition for which any required $\mathcal F_l^q(h)$ is empty is infeasible. The formulation assumes additive layer costs and does not model cross-layer overlap or inter-layer transfer beyond the simulator estimates.

\subsection{Candidate Paths and Search}
\textbf{MAC-guided pruning.} Similar to contraction-order search for general tensor expressions~\cite{smith2018opt}, a depth-first exploration builds contraction orders, eliminates equivalent candidates and high-MAC branches, and retains up to $K$ low-MAC paths per layer and mode. MAC count is used only to prune the space; the final choice is made by simulated latency. Independent contraction branches can be mapped concurrently to the two fixed subarrays in a feasible split configuration~\cite{tian2025ultra}.

\textbf{Latency-based selection.} SCALE-Sim v3~\cite{raj2025scale} evaluates each feasible $(p,h,d)$ for the corresponding layer and execution mode. For each $h$, the minimum estimated latency is found independently within every $(l,q)$ group and then added across layers and modes; the lowest combined total selects $h^*$. With at most $K$ paths, three partitions, and three dataflows, at most $2|\mathcal H||\mathcal D|KL=18KL$ mode/layer/configuration evaluations are required, and the selection is exact over the retained candidates.

\subsection{Hardware Realization}
\label{sec:hardware}
The selected mappings are executed on a configurable FPGA accelerator written in C++ HLS, which serves to validate the search results in hardware. A parameterizable $M_\mathrm{PE}\!\times\!N_\mathrm{PE}$ INT8 systolic GEMM engine, whose tiling unit maps the irregular $(M,K,N)$ GEMMs of TT contractions into $\langle T_M,T_N,T_K\rangle$ tiles, is driven by a streaming TT contraction kernel that issues each layer's contractions in the selected order. The array is organized as one array or as two $M_\mathrm{PE}\!\times\!\frac{N_\mathrm{PE}}{2}$ or $\frac{M_\mathrm{PE}}{2}\!\times\!N_\mathrm{PE}$ subarrays according to $h^*$, which stays fixed during execution; independent branches are assigned to the two subarrays when feasible, and dependent contractions are scheduled after their inputs become available. The engine uses routing multiplexers, configurable on-chip buffer roles, and PE control registers to select IS/OS/WS dataflows for the scheduled workload without changing the physical partition or regenerating the bitstream. Forward contractions and the backward computations of the evaluated TT linear and convolution layers are scheduled on the same engine, which supports both inference and training.

\section{Experimental Results}
\label{sec:experiments}

\subsection{Setup}
We evaluate ResNet-18 on CIFAR-10 and Tiny ImageNet and ViT-Ti/4 on CIFAR-10, for inference and training. TT ranks and INT8 precision are fixed before the search. As Table~\ref{tab:acc} shows, the TT+INT8 models have 38.72$\times$, 35.82$\times$, and 12.17$\times$ fewer parameters than the original models, at accuracy reductions of 0.71, 2.12, and 2.74 percentage points. The search keeps the TT ranks and the quantized model fixed, so accuracy is determined solely by the evaluated TT+INT8 representations.

\begin{table}[!t]
\centering
\caption{Top-1 accuracy and parameter compression.}
\label{tab:acc}
\setlength{\tabcolsep}{4pt}
\begin{tabular}{lccc}
\toprule
\textbf{Model / Dataset} & \textbf{Original (\%)} & \textbf{TT+INT8 (\%)} & \textbf{Params$\downarrow$} \\
\midrule
ResNet-18 / CIFAR-10      & 93.33 & 92.62 & 38.72$\times$ \\
ResNet-18 / Tiny ImageNet & 50.81 & 48.69 & 35.82$\times$ \\
ViT-Ti/4 / CIFAR-10       & 81.84 & 79.10 & 12.17$\times$ \\
\bottomrule
\end{tabular}
\end{table}

The simulator models $32\!\times\!32$ PEs with 3072\,KB input/weight and 1024\,KB output SRAM; the $1\!\times\!2$ and $2\!\times\!1$ partitions thus correspond to two $32\!\times\!16$ and two $16\!\times\!32$ subarrays, respectively, and up to $K=6$ paths are retained per layer and mode. The accelerator is implemented with Vitis HLS 2024.2 on a Xilinx VU9P at 200\,MHz. The \emph{Dense} baseline uses INT8 arithmetic under the same FPGA and PE-budget setting; \emph{TT+DSE} executes the TT+INT8 representation with the joint-mode, fixed-partition mapping.

\begin{figure}[!t]
\centering
\includegraphics[width=\linewidth]{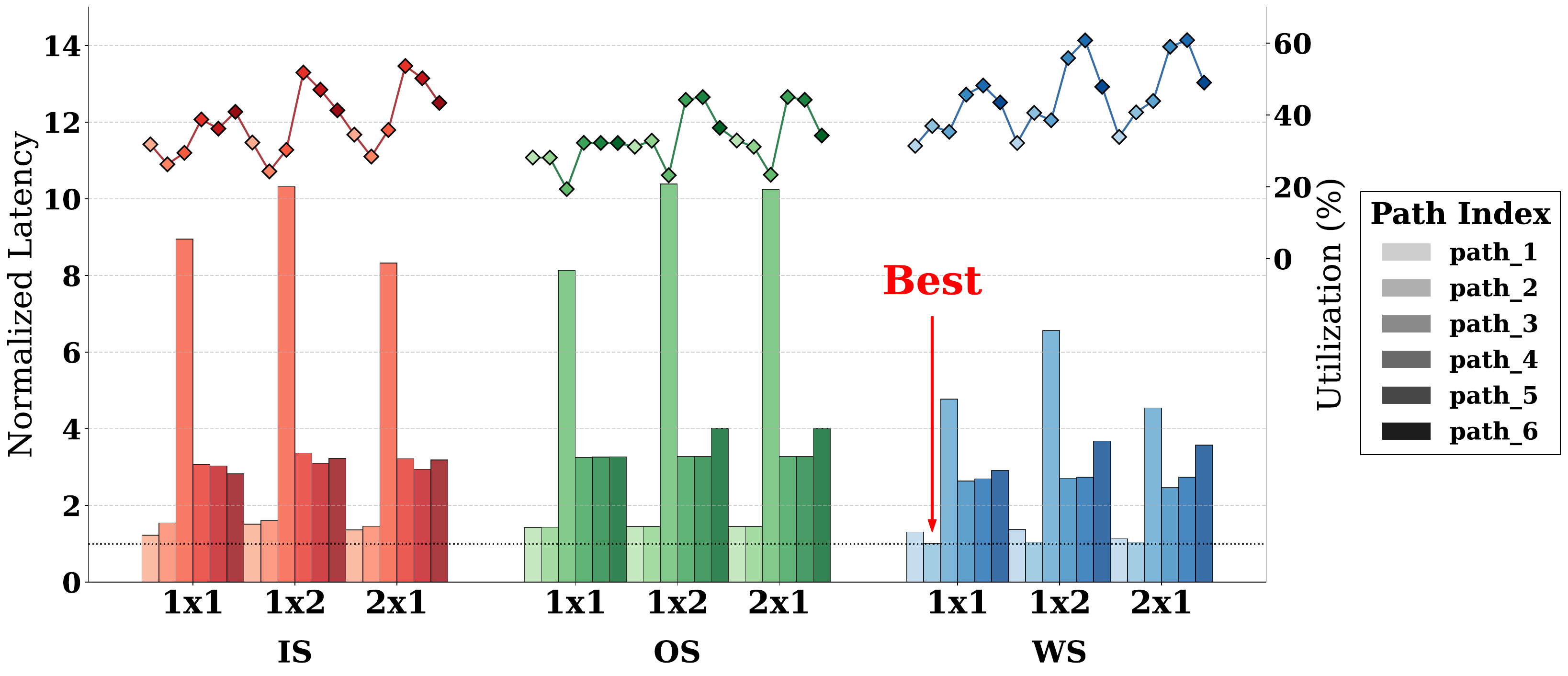}
\caption{Layer-level simulated latency of the retained contraction paths under three PE partitions and IS/OS/WS dataflows.}
\label{fig:layer}
\end{figure}

\begin{table}[!t]
\centering
\caption{Layer-level mapping statistics. \emph{Split pref.}: layers whose fastest configuration uses a split array if each layer could choose its own partition. \emph{Non-min-MAC}: layers whose selected path is not the minimum-MAC one. Last column: share of layers using each dataflow.}
\label{tab:decisions}
\setlength{\tabcolsep}{3pt}
\begin{tabular}{llccc}
\toprule
\textbf{Bench.} & \textbf{Mode} & \textbf{Split pref.} & \textbf{Non-min-MAC} & \textbf{IS / OS / WS (\%)} \\
\midrule
\multirow{2}{*}{RN18-C10} & Inf.  & 100\% & 25\% & 25 / 75 / 0 \\
                          & Train & 100\% & 25\% & 43.8 / 56.2 / 0 \\
\addlinespace[2pt]
\multirow{2}{*}{RN18-TIN} & Inf.  & 100\% & 25\% & 50 / 50 / 0 \\
                          & Train & 100\% & 25\% & 62.5 / 37.5 / 0 \\
\addlinespace[2pt]
\multirow{2}{*}{ViT-C10}  & Inf.  & 50\%  & 50\% & 0 / 0 / 100 \\
                          & Train & 50\%  & 50\% & 0 / 66.7 / 33.3 \\
\bottomrule
\end{tabular}
\end{table}

\begin{table}[!t]
\centering
\caption{FPGA results on VU9P (200\,MHz, same total PE count).}
\label{tab:fpga}
\setlength{\tabcolsep}{3.2pt}
\resizebox{\columnwidth}{!}{%
\begin{tabular}{llccccc}
\toprule
\textbf{Bench.} & \textbf{Method} & \textbf{BRAM} & \textbf{FF} & \textbf{LUT} & \textbf{Power (W)} & \textbf{Lat. (ms)} \\
\midrule
\multicolumn{7}{c}{\textit{Inference}}\\
\midrule
\multirow{2}{*}{RN18-C10} & Dense  & 892  & 342k & 318k & 14.2 & 6.40 \\
                          & TT+DSE & 956  & 408k & 390k & 11.8 & 1.60 (4.00$\times$) \\
\addlinespace[2pt]
\multirow{2}{*}{RN18-TIN} & Dense  & 1125 & 415k & 388k & 15.8 & 19.20 \\
                          & TT+DSE & 922  & 475k & 452k & 12.5 & 4.90 (3.92$\times$) \\
\addlinespace[2pt]
\multirow{2}{*}{ViT-C10}  & Dense  & 1085 & 368k & 345k & 15.1 & 8.69 \\
                          & TT+DSE & 1010 & 422k & 396k & 12.7 & 2.65 (3.28$\times$) \\
\midrule
\multicolumn{7}{c}{\textit{Training}}\\
\midrule
\multirow{2}{*}{RN18-C10} & Dense  & 1580 & 452k & 428k & 25.8 & 20.40 \\
                          & TT+DSE & 1482 & 524k & 525k & 21.2 & 5.30 (3.85$\times$) \\
\addlinespace[2pt]
\multirow{2}{*}{RN18-TIN} & Dense  & 1725 & 540k & 510k & 26.1 & 61.50 \\
                          & TT+DSE & 1428 & 595k & 565k & 22.8 & 16.10 (3.82$\times$) \\
\addlinespace[2pt]
\multirow{2}{*}{ViT-C10}  & Dense  & 1620 & 481k & 458k & 26.7 & 24.30 \\
                          & TT+DSE & 1515 & 531k & 534k & 22.5 & 7.11 (3.42$\times$) \\
\bottomrule
\end{tabular}}
\end{table}

\subsection{Layer-Level Mapping Analysis}
\textbf{Layer-level sensitivity.} Fig.~\ref{fig:layer} reports the simulated latency and utilization of retained paths under the three partitions and IS/OS/WS mappings. Paths are indexed in increasing MAC order, so path\_1 is the minimum-MAC candidate. For this layer, simulated latency spans more than $10\times$ across the retained configurations, and the fastest one uses path\_2 with WS on the single array, rather than path\_1. Different contraction shapes exhibit different utilization, so lower MAC count alone need not imply lower execution time, which motivates selecting mappings by simulated latency.

\textbf{Contraction paths.} Table~\ref{tab:decisions} shows that 25\% of ResNet-18 layers and 50\% of ViT-Ti/4 layers use a path other than the minimum-MAC candidate under the selected mapping. For the ViT layer in Fig.~\ref{fig:path}, the latency-selected path reduces the estimated cycle count from $1.36\times10^4$ to $1.04\times10^4$ (23.5\%) despite slightly more MACs.

\textbf{Dataflows.} ResNet-18 mappings alternate between IS and OS, whereas the ViT-Ti/4 inference mappings select WS and its training mappings predominantly select OS. No single preselected dataflow therefore fits every contraction, which motivates per-layer dataflow selection.

\textbf{Shared partition.} The split-preference column reports what each layer would prefer if allowed to choose its own physical partition: 100\% of evaluated ResNet-18 layers and 50\% of ViT-Ti/4 layers prefer a split. The global search resolves these competing layer-level preferences under a shared partition constraint; Eq.~\eqref{eq:obj} instead selects one partition by combining inference and training costs, while retaining the mode-specific path and dataflow choices reflected in the other columns. The selected partition is $h^*=2\!\times\!1$ (two $16\!\times\!32$ subarrays) for ResNet-18 and $h^*=1\!\times\!2$ (two $32\!\times\!16$ subarrays) for ViT-Ti/4; although only half of the ViT-Ti/4 layers prefer a split, the combined inference and training cost still favors one.

\subsection{FPGA Results}
Compared with Dense, TT+DSE reduces latency by 3.28$\times$--4.00$\times$ for inference and 3.42$\times$--3.85$\times$ for training (Table~\ref{tab:fpga}), and power by 15.9\%--20.9\% and 12.6\%--17.8\%. These ratios reflect TT compression together with the selected mapping. The added control and dataflow flexibility costs 11\%--23\% more LUTs. BRAM changes by only $-18.0\%$ to $+7.2\%$, far less than the parameter reduction, because BRAM also holds activations, intermediate tensors, and tile buffers, not only weights. For each model, the inference and training builds use the same partition $h^*$; the training build additionally contains the backward and weight-update logic with its buffers, which accounts for its higher BRAM, FF, and LUT usage, a trend also seen in the Dense baselines.

\section{Conclusion}
We presented a deployment-constrained, latency-guided DSE framework that selects a single physical PE partition shared by all layers and both inference and training, while choosing contraction paths and stationary dataflows per layer and mode. The layer-level results expose mismatches between MAC count and simulated latency, while the VU9P implementation demonstrates 3.28$\times$--4.00$\times$ inference and 3.42$\times$--3.85$\times$ training speedups over the INT8 dense baseline under the same PE budget. Future work will extend the deployment-constrained exploration to adaptive tensor ranks and larger-scale neural networks.

\clearpage
\bibliographystyle{IEEEtran}
\bibliography{references}
\end{document}